\documentclass[10pt]{article}

\usepackage{amsmath,xspace,bbm}
\usepackage{abstract}
\usepackage{amssymb}
\usepackage{amsthm}
\usepackage{mathtools}
\usepackage[dvips]{color}

\title{Rational conformal field theory with matrix level\\ and strings on a torus\\[6pt]\ }

\author{Ali Nassar \& Mark A. Walton\\[6pt]
Department of Physics and Astronomy\\  University of Lethbridge\\  Lethbridge,  Alberta, Canada\ \  T1K 3M4\\
\texttt{nassar@uleth.ca}, \texttt{walton@uleth.ca}
}

\begin{document}

\maketitle
\begin{abstract}
Study of the matrix-level affine algebra $U_{m,K}$ is motivated by conformal field theory and the fractional quantum
Hall effect. Gannon completed the classification of $U_{m,K}$ modular-invariant partition functions. Here we connect
the algebra $U_{2,K}$ to strings on 2-tori describable by rational conformal field theories. As Gukov and Vafa
proved, rationality selects the complex-multiplication tori. We point out that the rational conformal field theories
describing strings on complex-multiplication tori have characters and partition functions identical to those of
the matrix-level algebra $U_{m,K}$. This connection makes obvious that the rational theories are dense in the moduli space of strings on $T^m$, and  may prove useful in other ways.

%
\end{abstract}

\section{Introduction}

2-dimensional conformal field theories (CFTs) are quite well understood.  The local conformal symmetry, described by the Virasoro algebra and its extensions,
is in many cases powerful enough to give a complete determination of the operator
spectrum, as well as  explicit formulas for the correlation functions.
A particularly simple class are \textit{rational} conformal field theories (RCFTs), characterized by having a finite number of primary fields \cite{Cardy:1986ie}.
They are consistently described by a finite set of representations  of a certain chiral algebra. Moreover the corresponding  genus-1 characters  $\chi_i (q)$ form a finite dimensional unitary representation of the modular group $PSL(2,\mathbb{Z})$	
\begin{equation}
\chi_i\bigg(\frac{a\tau +b}{c \tau+d}\bigg)=\sum_j M_i^j \chi_j(\tau),
\end{equation}
where $a,b,c,d \in \mathbb{Z}$ and $ad-bc=1$.

One arena where RCFTs arise is in string compactifications on Calabi-Yau manifolds at special values of  their moduli, e.g.,  in Gepner models \cite{Gepner:1987qi}.  It is important to understand the conditions for rationality.  For example, the simplest compactifications of a string theory are on tori $T^m$--when are they described by RCFTs?  This question has been studied in \cite{Harvey:1987da,Wendland:2000ye,Moore:1998pn,Gukov:2002nw}. In particular, Wendland \cite{Wendland:2000ye} derived rationality conditions valid for all torus  dimensions $m$.  More importantly for us, in the case of 2-tori, Gukov and Vafa \cite{Gukov:2002nw} found a simple, geometric criterion for rationality. For $T^2$, the modular parameter $\tau$ and the K\"{a}hler parameter $\rho$ must take special values; they must belong to an imaginary quadratic number field. Such 2-tori have the property of complex multiplication, and are known as CM tori.  For them, a Gauss product exists, and was used in  \cite{Hosono:2002yb} to classify the corresponding RCFTs.

The special values of the moduli indicate RCFTs where the infinite number of fields is organized into a finite set that is primary with respect to an extended chiral algebra \cite{Moore:1989yh}. In the $T^m$ case, the generic boson algebra $U(1)^m= {U(1)}_{k_1}\times \cdots \times {U(1)}_{k_m}$ is extended by vertex operators defined by vectors of the lattice $\Lambda_m$ describing $T^m\cong {\mathbb R}^m/\Lambda_m$. The extended algebra is well understood; it was written explicitly for the $m=1$ case in \cite{Moore:1989yh}, for example.

On the other hand, a different algebra has also been of interest. The Abelian algebra  $U_{m,K}$ with an $m\times m$ matrix-valued level $K$ was studied in \cite{Gannon:1996hp} where its modular invariant partition functions were classified.
This matrix-level algebra generalizes $U(1)^m= {U(1)}_{k_1}\times \cdots \times {U(1)}_{k_m}$, which is recovered for a  diagonal matrix $K$. Here we consider the more general case,  allowing $K$ to be a non-negative integer-valued matrix.

Study of the matrix-level affine algebra $U_{m,K}$ is partly motivated by an effective description of the fractional quantum Hall effect, via a Chern-Simons theory with a gauge group $U(1)^m$  and a matrix valued level $K$ \cite{Belov:2005ze,Cappelli:1996np}. The  Witten correspondence \cite{Witten:1988hf, Moore:1989yh} relates the Chern-Simons theory with a gauge group $G$ canonically quantized on a manifold $M = \Sigma \times \mathbb{R}_t$, to the RCFT based on the  affine Kac-Moody algebra $\hat{G}_k$
(the Wess-Zumino-Witten model). For any simple, compact gauge group $G$ the Chern-Simons actions on a three-manifold  are classified by an integer $k \in \mathbb{Z}$.\footnote{If the three-manifold is a spin manifold then $k$ can be half-integer \cite{Dijkgraaf:1989pz}.} It was shown in \cite{Belov:2005ze} that Chern-Simons theories for the Abelian group $U(1)^m$ are classified by positive integer lattices with intersection forms $K$. Due to their potential application in the fractional quantum Hall effect, it is of interest to study the two-dimensional avatars of the  $U(1)^m$ Chern-Simons theories based on matrix valued level $K$. This reduces to the study of two-dimensional RCFTs with the affine Kac-Moody algebra $U_{m,K}$.

In this paper, we will uncover a connection between the matrix-level algebras $U_{2,K}$ and the CM tori, and so to the extended Moore-Seiberg algebras that make strings on them describable by RCFTs.
%
We point out that the RCFTs which arise from strings on CM-tori and the RCFTs based on the $U_{m,K}$  algebra have the same set of characters and the same partition function.

This relation has already proven useful in the following way.  As was noted in \cite{Gannon:1996hp},
the moduli space of the RCFTs based on the $U_{m,K}$  algebra is given in terms of the moduli space of rational points on the  Grassmannian $G_{d,d}(\mathbb{R})$. This is similar to the Narain moduli space of compactifications of strings on tori.  We  show that the characterization of the $U_{m,K}$ partition functions in terms of rational points on a Grassmannian \cite{Gannon:1996hp} is equivalent to specifying CM tori inside the Narain moduli space. This is another way to show that the set of RCFTs is dense in the Narain moduli space since the set of rational points is dense in the  Grassmannian $G_{d,d}(\mathbb{R})$. We hope that the relation between matrix-level RCFTs and strings on tori will also be helpful in other ways.

This paper is organized as follows.  In Section \ref{sec:2}, we discuss the $U_{m,K}$ matrix level affine algebras and their modular invariant classification \cite{Gannon:1996hp}.  In Section \ref{sec:3}, we study strings on CM tori and state the conditions for rationality, drawing from the results in \cite{Gukov:2002nw}. Using the Gauss product  (introduced into this context by \cite{Hosono:2002yb}), we relate the geometry of CM tori to the abstract $U_{m,K}$ algebras for $m=2$.
In Section \ref{sec:4}, we formulate the problem of RCFTs based on CM tori in terms of rational points on Grassmannians and as such we relate it the $U_{m,K}$ classification given in \cite{Gannon:1996hp}. Section \ref{sec:5} is a short conclusion.


\section{Matrix-level affine algebras}\label{sec:2}

In this section, we study affine Abelian chiral algebras with matrix-valued level. To provide evidence that they are the chiral algebras of a  consistent class of RCFTs, we describe their modular invariant partition functions. In Section {\ref{sec:3}} we relate them to $\sigma$-models on CM tori.


Let $\Gamma_K$ be a Euclidean, even, integral lattice of rank $r$ with a positive definite integer-valued symmetric
 intersection matrix $K_{ij}$, of determinant $-D>0$.
The basis $\{\mathbf{e}_i\}$ of  $\Gamma_K$ is defined up to $GL(r,\mathbb{Z})$  transformations which preserve the determinant of $K$
\begin{equation}
 \mathbf{e}_i \rightarrow G_i^j  \mathbf{e}_j, \quad G \in  GL(r,\mathbb{Z})\ ,\  \quad |G|=\pm 1\, .
\end{equation}
where $|G|$ is the determinant of $G$.

We can associate with any integral lattice $\Gamma_K$ of rank $r$ a chiral vertex algebra $\mathcal{A}(\Gamma_K)$. There are $r$ linearly independent $U(1)$ currents 
\begin{equation}
J_i(z)=\sum_{n\in \mathbb{Z}} \frac{J^n_i}{z^{n+1}}.
\end{equation}
They have the OPE
\begin{equation}
J_i(z)J_j(w) \sim\frac{K_{ij}}{(z-w)^2}.
\end{equation}
In terms of the modes we have
\begin{equation}
\big[J^m_i,J^n_j\big] = m K_{ij} \delta_{m+n,0}.
\end{equation}
These spin-one currents form a subalgebra $\mathcal{A}_r \subset \mathcal{A}(\Gamma)$  with rank $r$.
The energy momentum tensor is given by the Sugawara construction
\begin{equation}
T(z)=\frac{1}{2} K^{ij}  : J_{i}(z) J_{j}(z): ,
\end{equation}
where $K^{ij}=K_{ij}^{-1}$ is the intersection form of the dual lattice.

From the OPE of $T(z)$ with itself we can read off the central charge
\begin{equation}
c=K^{ij} K_{ij}=r.
\end{equation}
The Virasoro generators are given by
\begin{equation}
L_n =\frac{1}{2} K^{ij} \sum_{m=-\infty}^{\infty}: J_i^{m+n} J_j^{-m} :.
\end{equation}

Now we specialize to rank two lattices $\Gamma_K$ with basis  $\mathbf{e}_1, \mathbf{e}_2$. Since the lattice is even-integer lattice, then its intersection form can be written as
\begin{equation}
K_{ij}=\langle\mathbf{e}_i|\mathbf{e}_j \rangle=
\begin{pmatrix}
2a &b\\
b &2c\\
\end{pmatrix}, \quad a,b,c\in\mathbb{Z}.
\end{equation}
We assume that $\text{gcd}(a,b,c)=1$ which corresponds to primitive quadratic forms.

The $GL(2,\mathbb{Z})$ transformation on the  basis of  $\Gamma_K$ gives an equivalent lattice
\begin{equation}
K\longrightarrow K'= G^t K G,\qquad \Gamma_{K'}\equiv \Gamma_K, \quad G\in GL(2,\mathbb{Z}).
\end{equation}
Since the determinant of $K$ is invariant under this transformation, we define
the set of equivalence classes of primitive, even lattices  as
\begin{equation}\label{glz}
\mathcal{L}^p(D):=\big\{\Gamma_K :  D = -| K|\big\} / GL(2,\mathbb{Z}).
\end{equation}
where  $|K|$ is the determinant of $K$.

We will consider different matrix levels $K_L$ for the holomorphic  and $K_R$ for the anti-holomorphic sectors which give rise to heterotic theories. The set of standard representations of the affine algebra $U_{m,K}$ are labeled by
$a\in P_+^{K_L} =\Gamma_{K_{L}}^*/\Gamma_{K_{L}} $ and $b\in P_+^{K_R} =\Gamma_{K_{R}}^*/\Gamma_{K_{R}} $ where $|K_L|=|K_R|$ \cite{Gannon:1996hp}.

The RCFT data are given by $(\Gamma_{K_{L}}, \Gamma_{K_{R}},\{\chi_a^{\Gamma_{K_L}}\}, \{\chi_a^{\Gamma_{K_R}}\})$ where $\chi_a^{\Gamma_{K_R}}$ are the characters which are proportional to the theta functions of the lattice
\begin{equation}\label{theta1}
\chi_a^{\Gamma_{K_L}}(q)=\frac{\theta_a^{\Gamma_{K_L}}(q)}{\eta(q)^2}=\frac{1}{{\eta(q)^2}}\sum_{v\in\Gamma_{K_L}} q^{\frac{1}{2}(a+v)^2},
\end{equation}
where $\eta(q)$ is the Dedekind eta function.



The spectrum is encoded in the  (genus-1) partition function
\begin{equation}\label{partitionfunction}
Z^{\Gamma_{K_{L}},\Gamma_{K_{R}}}=\sum_{a\in P_+^{K_L},\; \;  b\in P_+^{K_R}} M_{a,b} \; \chi_a^{\Gamma_{K_{L}}} \overline{ \chi_{b}^{\Gamma_{K_{R}}}},
\end{equation}
where the matrix $M_{a,b}$ is constrained to satisfy $M_{a,b}\in \mathbb{Z}_{\geq}$ and  $M_{0,0}=1$.

Modular invariance dictates that
\begin{equation}\label{ST}
SM=MS,\qquad TM=MT,
\end{equation}
where the matrices  $S$ and $T$ are unitary and symmetric and $T$ is diagonal
\begin{equation}
S_{ab}=\frac{1}{\sqrt{|K_L|}}\exp[-2\pi i (a\cdot b)].
\end{equation}

In \cite{Gannon:1996hp}, all modular invariants of the algebra $U_{m,K}$ were constructed in terms of even self-dual lattices $\Gamma$ that contain $\Gamma_{K_{L}}$ and $\Gamma_{K_{R}}$.
In terms of the matrix $K$, full modular invariance means  $K_{ij}\in 2 \mathbb{Z}$.  This, together with the symmetry of $K$, translates to an even integer lattice $\Gamma_K$.

The argument in \cite{Gannon:1996hp} goes as follows: consider the heterotic partition function (\ref{partitionfunction}).
 Using $SM=MS$ and the unitarity of the $S$ matrix we find
\begin{equation}
\begin{split}
M_{a,b} &= \sum_{c,d} S_{a,c} M_{c,d} S_{d,b}^*\\
&=\frac{1}{|K_L|}\sum_{c,d} \exp[2\pi i (b\cdot d-a\cdot c ) ] M_{c,d}.
\end{split}
\end{equation}
From this equation we derive the relation
\begin{equation}\label{eq:gannon}
\frac{M_{a,b}}{ M_{0,0}}=\frac{\sum_{c,d} z_{ab,cd} M_{c,d}}{\sum_{c,d}  M_{c,d}}. 
\end{equation}
where we defined
\begin{equation}
 z_{ab,cd}=\exp[2\pi i (b\cdot d-a\cdot c ) ], \quad | z_{ab,cd}|=1.
\end{equation}
Using the triangle inequality, the above equation gives $|M_{a,b}|\leq |M_{0,0}|=1$ with the equality iff
\begin{equation}
M_{c,d}\neq 0 \Longrightarrow b\cdot d =a\cdot c \quad (\text{mod }1)
\end{equation}
for all $c \in P_+^{K_L}, d\in P_+^{K_R}$. Hence $M_{a,b}=\{0,1\}$. 
 Define the set
\begin{equation}
\Omega=\bigcup_{a\in  P_+^{K_L},\; b\in  P_+^{K_R} } ( a\oplus ib)+ \big(\Gamma_{K_{L}} \oplus i\Gamma_{K_{R}}\big)
\end{equation}
which is an even self dual lattice. The matrix $M_{a,b}$ in (\ref{partitionfunction}) satisfies
\begin{equation}\label{Mab}
M_{a,b}=
\begin{cases}
1 \quad \text{if } (a,i b) \in \Omega \\
0 \quad \text{otherwise}.
\end{cases}
\end{equation}
Hence,  the modular invariant partition functions of the $U_{m,K}$ current algebra are in 1-to-1 correspondence with the even self-dual lattices $\Gamma$ which contain $(\Gamma_{K_{L}};\Gamma_{K_{R}})$.





\section{Strings on a CM torus}\label{sec:3}

Consider an elliptic curve (or a torus)
 $E_\tau=\mathbb{C}/\Lambda $, where $\Lambda$ is a lattice  $\Lambda=(\mathbb{Z}+\tau\mathbb{Z})$ and $\tau$ is the complex structure modulus.
The endomorphisms of $E_\tau$ are given by holomorphic maps $ F: E_\tau\rightarrow E_\tau$.
  Clearly any elliptic curve has trivial endomorphisms corresponding to multiplication by an integer $\lambda \in \mathbb{Z}$. We want to find the conditions on $\tau$ which give non-trivial endomorphisms. The action of $\lambda$ on the lattice $\Lambda$ is represented on the generators as \cite{Moore:1998pn}
\begin{equation}
\begin{split}
\lambda\cdot 1&= m_1 \cdot 1 +n_1 \cdot \tau \\
\lambda\cdot \tau &= m_2 \cdot 1 +n_2 \cdot \tau ,
\end{split}
\end{equation}
where $m_1,m_2,n_1,n_2\in \mathbb{Z}$. By substituting the first equation in the second  we learn that for an elliptic curve to have non-trivial endomorphisms, then $\tau$ must satisfy a quadratic equation with integer coefficients
\begin{equation}\label{tau-equation}
a\tau^2+b\tau+c=0,\quad \tau =\frac{-b+ \sqrt{D}}{2a},\ \ D=b^2-4ac < 0.
\end{equation}
This means that $\tau$ takes values in the imaginary quadratic number field $\mathbb{Q}(D)$,  the set of numbers of the form $\alpha+\beta \sqrt{D}$ with $\alpha, \beta \in \mathbb{Q}$.

Elliptic curves (or tori)  for these special values of $\tau$ are said to have \textit{complex multiplication} (or to be of CM type).  The $\sigma$-model on $E_\tau$ is specified by another parameter, the  K\"ahler parameter  $\rho$.
The compactification of strings on  $E_\tau$  is characterized by a momentum-winding Narain lattice, an even self-dual lattice $\Gamma(\tau,\rho)$ of rank $4$,  where the parameters $\tau$ and $\rho$  live in the upper-half  plane $\mathbb{H}^+$ subject to a group of discrete symmetries $\Xi$. The Narain moduli space of this compactification is \cite{Gukov:2002nw}
\begin{equation}\label{2d-narain-moduli}
\mathcal{M}=\frac{\mathbb{H}^{+}\times \mathbb{H}^{+}}{\Xi},
\end{equation}
where $\Xi$ is the group of discrete symmetries of $E_\tau$ 
\begin{equation}
\Xi=PSL(2,\mathbb{Z})_\tau \times PSL(2,\mathbb{Z})_\rho \times \mathbb{Z}_2 \times \mathbb{Z}_2 \times \mathbb{Z}_2.
\end{equation}
The first  $\mathbb{Z}_2$ is a mirror symmetry which exchanges $\tau$ and $\rho$, i.e., $\mathbb{Z}_2: (\tau,\rho)\mapsto (\rho,\tau)$. The second   $\mathbb{Z}_2$ is a space-time parity transformation $\mathbb{Z}_2: (\tau,\rho)\mapsto (-\bar{\tau},-\bar{\rho})$, where the bar denotes complex conjugation. The last $\mathbb{Z}_2$  is a world-sheet orientation reversal $\mathbb{Z}_2: (\tau,\rho)\mapsto (\tau,-\bar{\rho})$ \footnote{The three $\mathbb{Z}_2$ symmetries are not really independent.}.

For generic values of $\tau$ and $\rho$, the conformal field theory is not rational and has an infinite number of primary fields.
For special values of $\tau$ and $\rho$  the infinite number of primary-field representations reorganize  into a finite set of representations of a bigger chiral algebra and the theory becomes rational.
It was shown in \cite{Gukov:2002nw} that CFTs based on $E_\tau$  are rational iff $E_\tau$ and its mirror are both of CM type, that is,  $\tau,  \rho\in \mathbb{Q}(D)$ which implies that $\rho$, like $\tau$, satisfies a quadratic equation with integer coefficients.

An example of an RCFT which enjoys this property results when $\tau=\rho=e^{2\pi i/3}$. In this case the chiral vertex operator algebra is isomorphic to
 $SU(3)$ WZW model at level $1$ \cite{Vafa:1989ih}. Clearly, both $\tau$ and $\rho$ satisfy
 \begin{equation}
 \tau^2+\tau+1=0.
\end{equation}

In \cite{Gukov:2002nw},  the diagonal case was studied in detail where it was shown that the condition for a  diagonal modular invariant is
\begin{equation}
\tau=f a\rho,
\end{equation}
where $f\in \mathbb{Z}$, {$a$ is the coefficient of $\tau^2$ in (\ref{tau-equation}), and $\tau,\rho \in \mathbb{Q}(D)$. The generalization to non-diagonal modular invariants was given in \cite{Hosono:2002yb}.

We can associate a quadratic form with the complex numbers $\tau$ and $\rho$. Write
\begin{equation}
Q(a,b,c)=\begin{pmatrix}
2a &b\\
b&2c
\end{pmatrix}, \qquad \tau_{Q(a,b,c)}=\frac{-b+\sqrt{D}}{2a}
\end{equation}
where $D=b^2-4ac=-|Q|$ denotes the discriminant of the quadratic form $Q$.  $Q$ is called primitive if $\text{gcd}(a,b,c)=1$. The discriminant of the quadratic form $Q$ is  invariant under $SL(2,\mathbb{Z})$:
\begin{equation}\label{class}
Q \longrightarrow S^t Q S, \quad S\in SL(2,\mathbb{Z}).
\end{equation}

Now we can consider  equivalence classes of quadratic forms under the action of $SL(2,\mathbb{Z})$ (or more precisely $PSL(2,\mathbb{Z})$, since $S=\pm \mathbb{I}$ acts trivially).  The set of equivalence classes is denoted by
\begin{equation}\label{ClD}
\begin{split}
Cl(D) = 
  \big\{Q (a,b,c)\, \big|\, D=b^2-4ac <0,\; a>0  \big\} / \sim SL(2,\mathbb{Z}).
\end{split}
\end{equation}
It is known that  $Cl(D)$ is a finite set  and we will denote the number of its elements by $h(D)$ (see \cite{Hosono:2002yb} and references therein)
\begin{equation}
Cl(D)=\big\{\mathcal{C}_1,\dots,\mathcal{C}_{h(D)}   \big\}.
\end{equation}
Since $D<0$, the complex number $\tau_Q$  lies  in the upper-half plane $\mathbb{H}^{+}$.
The $SL(2,\mathbb{Z})$ action on $Q$ will induce a fractional linear transformation on $\tau_Q$.
The $PSL(2,\mathbb{Z})$ orbits of  $\tau_{Q(a,b,c)} \in \mathbb{H}^{+}$ depend on the class $\mathcal{C}=[Q(a,b,c)]\in Cl(D)$.
Using the above mapping, then we can label the classes in $Cl(D)$ by points $[\tau_Q] \in \mathcal{F}=\mathbb{H}^{+}/PSL(2,\mathbb{Z})$.  Since $\tau$ is the complex structure of a torus then fractional linear transformations on $\tau$ gives an equivalent torus. The classes $[\tau_Q]$ will give inequivalent tori in the Narain moduli space.

The same goes for $\rho$ where we can also define equivalence classes of quadratic forms parametrized by points $[\rho_{Q'}] \in \mathcal{F}=\mathbb{H}^{+}/PSL(2,\mathbb{Z})$. The equivalence class of Narain lattices corresponding to an RCFT will be denoted by  $\Gamma(\tau_\mathcal{C},\rho_\mathcal{C'})$, where $\mathcal{C}$ and $\mathcal{C'}$ are the equivalence classes corresponding to $\tau$ and $\rho$.

Similarly, the group $GL(2,\mathbb{Z})$ acts on quadratic forms by the same formula as (\ref{class}). The set of \textit{improper} equivalence classes under $GL(2,\mathbb{Z})$ is defined  by \cite{Hosono:2002yb}
\begin{equation}
\begin{split}
\tilde{Cl}(D)=
 \big\{Q (a,b,c)\, \big|\, D=b^2-4ac <0,\; a>0  \big\} / \sim GL(2,\mathbb{Z}).
\end{split}
\end{equation}
and we have a surjection  $q$
\begin{equation}
q: Cl(D)\rightarrow \tilde{Cl}(D), \quad \mathcal{C}\rightarrow \tilde{\mathcal{C}},
\end{equation}
where $q^{-1}(\tilde{\mathcal{C}})$ has either one or two classes.

One can associate with the lattice $\Gamma_K$ a primitive quadratic form given by the intersection form $K$. Therefore, the equivalence classes of primitive lattices $[\Gamma_K]$ are in 1-to-1 correspondence with the equivalence classes of primitive, quadratic forms $[Q(a,b,c)]$. Hence,  we can identify the set $\tilde{Cl}(D)$ with the set $\mathcal{L}^p(D)$ in (\ref{glz}).

 We define the following projections of  $\Gamma(\tau_{\mathcal{C}},\rho_{\mathcal{C}'})$
\begin{equation}
\Pi_L :=\Gamma(\tau_{\mathcal{C}},\rho_{\mathcal{C}'})\cap \mathbb{R}^{2,0}, \qquad \Pi_R :=\Gamma(\tau_{\mathcal{C}},\rho_{\mathcal{C}'})\cap \mathbb{R}^{0,2},
\end{equation}
which correspond to the equivalence classes of the left and right momentum lattices,  characterized by  the vanishing of the right moving and left moving momenta, respectively.

The modular invariant partition functions studied in \cite{Hosono:2002yb} take the form:
\begin{equation}\label{automorphisms}
\begin{split}
Z^{\Pi_L,\Pi_R,\varphi}(q,\bar{q}) &=\frac{1}{|\eta(q)|^4} \sum_{a\in \Pi_L^*/\Pi_L} \theta_a^{\Pi_L}(q), \overline{\theta_{\varphi(a)}^{\Pi_R}(q)}\\[3pt]
& =\sum_{a\in \Pi_L^*/\Pi_L} \chi_a^{\Pi_L}(q), \overline{\chi_{\varphi(a)}^{\Pi_R}(q)}
\end{split}
\end{equation}
where $\varphi$ is a gluing map between the discriminant groups $\Pi_L^*/\Pi_L$ and $\Pi_R^*/\Pi_R$. It satisfies $(\varphi(a),\varphi(b))=(a,b)$, where $a,b\in \Pi_L^*/\Pi_L$ and $(\cdot,\cdot)$ is the rational bilinear form on $\Pi_L^*/\Pi_L$ which is induced from the bilinear form on $\Pi_L$.

The characters which enter the partition function above are given in terms of the  theta function of the lattice $\Pi_L $:
\begin{equation}\label{theta}
\chi_a^{\Pi_L}(q)=\frac{\theta_a^{\Pi_L}(q)}{\eta(q)^2}=\frac{1}{\eta(q)^2}\sum_{v\in\Pi_L} q^{\frac{1}{2}(a+v)^2}
\end{equation}
They are identical to the characters in (\ref{theta1}).

\subsection*{The Gauss product}

There is a binary operation which turns the set $Cl(D)$ of (\ref{ClD}) into an Abelian group:  the Gauss product (see \cite{Hosono:2002yb}, e.g.) takes two equivalence classes of quadratic forms of the same discriminant and produces a third with that discriminant.

Let  $\mathcal{C}=[Q_1(a_1,b_1,c_1)]$ and $\mathcal{C}'=[Q_2(a_2,b_2,c_2)]$ be two such equivalence classes. We will restrict ourselves to primitive forms. We say that two quadratic forms $Q_1(a_1,b_1,c_1)\in \mathcal{C}$ and $Q_2(a_2,b_2,c_2)\in \mathcal{C}'$ are concordant if $a_1 a_2 \neq 0$,  $\text{gcd}(a_1,a_2)=1$ and $b_1=b_2$. Then the Gauss product of  $\mathcal{C}\star\mathcal{C}'$ is defined as
\begin{equation}
\big[Q_1\big(a_1,b,c_1\big)\big] \star\big[Q_2\big(a_2,b,c_2\big)\big]=\big[Q_3\big(a_3,b_3,c_3\big)\big],
\end{equation}
where $a_3=a_1 a_2$, $b_3=b$, and $c_3=\frac{b^2-D}{4a_1 a_2}$.
It is important to mention that any pair of quadratic forms of the same discriminant can be $SL(2,\mathbb{Z})$-transformed to a concordant pair.

The unit $\mathbf{1}_D$ of $Cl(D)$ with respect to the product $\star$ is represented by
\begin{equation}
\mathbf{1}_D=
\begin{cases}
[1,0,-\frac{D}{4}]; & \text{if } D\equiv 0 \mod 4\\[5pt]
[1,0,\frac{1-D}{4}]; & \text{if } D\equiv 1 \mod 4.
\end{cases}
\end{equation}

The  quadratic form $Q_3(a_3,b_3,c_3)$ corresponds to a lattice with intersection form
\begin{equation}
Q_3\big(a_3,b_3,c_3\big)\equiv \begin{pmatrix}
2a_3 &b_3\\
b_3&2c_3
\end{pmatrix}.
\end{equation}
which gives an even integer lattice, an important fact which we will use  when we construct the matrix levels $K_L$ and $K_R$ for the $U_{m,K}$ algebras.

The Gauss product is used to construct the intersection form
of the lattices $\Pi_L$ and $\Pi_R$ in terms of the equivalence classes $\mathcal{C}$ and $\mathcal{C}'$,  corresponding to $\tau_{Q(a_1,b_1,c_1)}$ and $\rho_{Q(a_2,b_2,c_2)}$, respectively   \cite{Hosono:2002yb}:
\begin{equation}
\Pi_L=q(\mathcal{C}\star\mathcal{C'}^{-1}), \qquad \Pi_R=q(\mathcal{C}\star\mathcal{C'})(-1).
\end{equation}
Here $q$ is the natural map $Cl(D) \rightarrow \tilde{Cl}(D)$ and \linebreak $q(\mathcal{C}\star\mathcal{C'})(-1)$ means we multiply the quadratic form $q(\mathcal{C}\star\mathcal{C'})$ by $-1$.

As was shown in \cite{Hosono:2002yb}, to prove the above result one first constructs the $\mathbb{Z}$-basis for $\Pi_R$ of $\Gamma(\tau_{\mathcal{C}},\rho_{\mathcal{C}'})$ in terms of the equivalence classes of $\mathcal{C}$ and $\mathcal{C}'$ and then compares the resulting quadratic form of $\Pi_R$ with ${Q}_3(a_3,b_3,c_3)$ and similarly for $\Pi_L$.




Now we have geometric data represented by the rational Narain lattice $\Gamma(\tau_\mathcal{C},\rho_\mathcal{C'})$ which depends on the $\sigma$-model parameters $\tau$ and $\rho$ (both $\tau$ and $\rho$ are attached to an equivalence class of quadratic forms) and  algebraic, RCFT data $(\Gamma_{K_{L}}, \Gamma_{K_{R}},\{\chi_a\})$. The Gauss product can used to relate them in the following way. First we will look at $K_L$ and $K_R$ as quadratic forms and hence we can talk about their  respective equivalence classes under the $SL(2,\mathbb{Z})$ action, as we did with $Q$.  Now, a rational point in the Narain moduli space specified by the special values $\tau_{\mathcal{C}}$ and $\rho_{\mathcal{C}'}$ defines two equivalence classes of quadratic forms $\mathcal{C}$ and $\mathcal{C}'$. The mapping between the two sets of data is
\begin{equation}\label{gaussmapping}
K_L = q\big(\mathcal{C}\star C'^{-1}\big), \qquad K_R = q\big(\mathcal{C}\star\mathcal{C}'\big)(-1).
\end{equation}
We will need to apply  a symmetrisation map (half the sum of
the matrix and its transpose) to $K_L$ and $K_R$ if they are not symmetric or $SL(2,\mathbb{Z})$-equivalent to their symmetric forms.
This mapping  shows that the algebras $U_{m,K}$ can be given a geometric significance, by relating them to $\sigma$-models on CM tori.
To justify this, we notice that the characters of the RCFTs in (\ref{theta})  which are proportional to the theta functions of the momentum-winding lattice are the same as the characters (\ref{theta1})  of the algebra $U_{2,K}$. Both sets of characters are based on lattices which are constructed from the same set of geometric data using the Gauss product.
Also, the modular invariant partition functions (\ref{automorphisms}) are a subset of the modular invariant partition functions (\ref{partitionfunction}) of the algebra $U_{2,K}$ for which
\begin{equation}
M_{a,b} = \delta_{b,\phi(a)}.
\end{equation}


We note that the above mapping is not 1-to-1.  We can't start from $K_L$ and $K_R$ and construct  unique equivalence classes for $\tau$ and $\rho$. The matrices $K_L$ and $K_R$ are  representatives of  equivalence classes in the set $Cl(D)$ which have a finite number of elements. The algebraic description of the $U_{m,K}$ algebras depends only on $K_L$ and $K_R$ and as such is the same for all members of the set $Cl(D)$. On the other hand there is a geometric description for each member of the set  $Cl(D)$. We conclude that the same $U_{m,K}$ algebra have many geometric avatars. This is similar to the case of a rational boson on a circle of radius square $R^2=p/q$ which is described by a level $k=pq$  $U(1)_k$ algebra.  On the other hand, starting from the algebra $U(1)_k$, there are many candidate rational boson theories, one for each  factorization of $k$ into two coprime integers $k=pq$.

\section{Rational points on Grassmannians and CM tori}\label{sec:4}

In this section we study the rationality conditions of a Narain lattice in more detail. We  formulate the rationality in terms of rational points on a Grassmannian and we show that these points are equivalent to tori of CM type. Our argument will be based on the results in \cite{Wendland:2000ye}.

We consider a generic Narain lattice $\Gamma(\tau,\rho)=(P_L;P_R)$ of the $\sigma$-model on $T^d/\Lambda$ with a $B$-field.
 The holomorphic and anti-holomorphic vertex operators are characterized by $P_R=0$ and $P_L=0$ and they are parametrized by the values of their charges in $\Pi_{L}=(P_L;0)$ and $\Pi_{R}=(0;P_R)$.
We also define the following projections of the lattice $\Gamma(\tau,\rho)$:
$\widetilde{\Pi}_{L}=(P_L;*)$  and
$\widetilde{\Pi}_{R}=(*;P_R) $
where the $*$ means we forget about the corresponding component of $P$.

Note that
\begin{equation}
\Pi_{L}\subseteq \widetilde{\Pi}_{L}, \qquad \Pi_{R}\subseteq \widetilde{\Pi}_{R}.
\end{equation}
Since the lattice $\Gamma(\tau,\rho)$ is even, self-dual and integral then its straightforward to show that
\begin{equation}
\Pi_{L}^* \cong \widetilde{\Pi}_{L}, \qquad \Pi_{R}^* \cong \widetilde{\Pi}_{R},
\end{equation}
i.e., $\widetilde{\Pi}_{L}$ is the dual of $\Pi_{L}$ and the same for $\widetilde{\Pi}_{R}$.

Rationality can be expressed in terms of the rank of $\Pi_L$ and $\Pi_R$ .
The Narain lattice $\Gamma(\tau,\rho)$ is rational if and only if \cite{Wendland:2000ye}
\begin{equation}\label{rankVSrationality}
\text{rank} \big(\Pi_L\big) =\text{rank} \big(\Pi_R\big)=d.
\end{equation}

RCFTs are characterized by the appearance of extra holomorphic vertex operators which extend the chiral algebra. For a generic Narain lattice $\Gamma(\tau,\rho)$, the only holomorphic vertex operator is the one corresponding to the vacuum $P_L=P_R=0$.
Since any field is mutually local
to the vacuum, then the set of allowed representations $\mathcal{V}=\{V[P], P \in \Gamma(\tau,\rho) \}$ is infinite.
However  extra holomorphic vertex operators $\mathcal{W}[P]$ appear for  $p\in \Pi_{L}$. The requirement of locality with respect to $\mathcal{W}[P]$ restricts the set of allowed irreducible representations to be $a\in \Pi_{L}^*/\Pi_{L}$, where $\Pi_{L}^*$ is the lattice dual to $\Pi_{L}$ so it contains charges which have integer product with $\Pi_{L}$. The condition for rationality translates to the requirement that $\Pi_{L}$ be a finite index sublattice of $\Pi_{L}^*$ so that the set $\Pi_{L}^*/\Pi_{L}$ has a finite cardinality. This happens when $\Pi_{L}$ have a finite rank which is the condition in (\ref{rankVSrationality}).

The Narain moduli space of conformal field theories is isomorphic to the moduli space of even, self-dual lattices with signature $(d,d)$
\begin{equation}\label{Grassmannian}
\mathcal{M}_d=O(\Gamma^{d,d})\backslash O(d,d)/(O(d)\times O(d)),
\end{equation}
where $\Gamma^{d,d}$ denotes the standard even self-dual lattice of signature $(d,d)$ and $O(\Gamma^{d,d})$ its automorphism group. In the special case of $d=2$ this gives the moduli space in (\ref{2d-narain-moduli}).
The above moduli space is also the Grassmannian of space-like $d$-planes in $\mathbb{R}^{d,d}$.
The modular invariant partition functions in \cite{Gannon:1996hp} are classified using rational points on the Grassmannian (\ref{Grassmannian}).

Note that
\begin{equation}
P_L\in W=\Pi_{L} \otimes \mathbb{R} , \qquad P_R \in  W^{\perp}= \Pi_R \otimes \mathbb{R},
\end{equation}
where $W$ is a space-like $d$-plane which correspond to a point on the Grassmannian (\ref{Grassmannian}) and $W^\perp$ is its orthogonal complement in $\mathbb{R}^{d,d}$.

The sets $\Pi_{L}$ and $\Pi_{R}$ in general are not lattices, since the set of vectors which span $\Pi_L$ and $\Pi_R$ will not remain linearly independent over $\mathbb{Z}$ when restricted to $W$ and $W^{\perp}$. However, if $W$ is a \textit{rational} point on the Grassmannian (\ref{Grassmannian})  then $\Pi_L$ and $\Pi_R$ become lattices of rank $d$. A rational point on the Grassmannian (\ref{Grassmannian}) is a subspace $W$ with basis $\{f_m\}$ which can be written over $\mathbb{Q}$ in terms of the preferred orthonormal basis $e_i$ of $\mathbb{R}^{d,d}$
\begin{equation}
f_m=Q_{mi} e_i, \quad Q_{mi}\in \mathbb{Q}.
\end{equation}
 The above equation implies that the basis of $W$ are rational vectors
 \begin{equation}
\langle  f_m| f_n\rangle\in \mathbb{Q}.
 \end{equation}
Since any group generated by rational
vectors is a lattice, i.e., it can be generated by linearly independent vectors over $\mathbb{Z}$. Then $\Pi_L$ which is the $\mathbb{Z}$-span of $f_m$ is a lattice of rank $d=\text{dim}(W)$ and the same for $\Pi_R$.  It was shown in \cite{Wendland:2000ye} that for $d=2$
\begin{equation}\label{rankcondition}
\text{rank} \big(\Pi_{L}\big) =\text{rank} \big(\Pi_R \big)=2 \longleftrightarrow \tau, \rho \in \mathbb{Q}(D).
\end{equation}
which in our case implies that rational points on the Grassmannian (\ref{Grassmannian}) correspond to CM tori.
This is another way to see that  RCFTs constitute a dense subset of the set of CFTs, since the set of rational points on a Grassmannian is dense which is consistent with the findings in \cite{Gukov:2002nw} that the values $\tau,\rho\in \mathbb{Q}(D)$ corresponding to RCFTs are dense in  in the Narain moduli space.

\section{Conclusion}\label{sec:5}

In summary, we studied the RCFTs based on the  matrix-level algebra $U_{2,K}$  and we related them to  strings on CM tori (corresponding to  $\tau,\rho \in \mathbb{Q}(D)$) inside the Narain moduli space. The characters and modular-invariant partition functions were shown to be identical in the 2 types of theories.\footnote{For characters, compare (\ref{theta}) to (\ref{theta1}); for partition functions, (\ref{automorphisms}) to (\ref{partitionfunction},\ref{Mab}).} Furthermore, the map between them was  constructed explicitly: the Gauss product was used to write the matrix levels $K_L$ and $K_R$ in terms of the geometric data represented by $\tau$ and $\rho$.

The connection was shown to be useful in 1 way. By formulating the problem in terms of rational points on a Grassmannian we showed that the set of RCFTs is a dense subset in the Narain moduli space. This agrees with the observation in \cite{Gukov:2002nw}  that the values of $\tau,\rho \in \mathbb{Q}(D)$ which produce RCFTs are dense in the Narain moduli space.
It is our hope that the relation between matrix-level algebras and RCFTs for strings on tori will prove useful in other ways.

It would be interesting to see if the results described here might be generalized to higher dimensions $m>2$, or to non-Abelian theories. Most exciting, perhaps, might be an explanation of the mysterious connection between bosons on a CM torus and matrix level. Does the  chiral  algebra of the former, extended as it is by vertex operators, have any more direct relation to the (simpler) Abelian matrix-level algebras, e.g.?

\bibliographystyle{plain}

\bibliography{preprint2}

\end{document}